\documentclass[english]{article}
\usepackage{authblk}
\usepackage{eucal}
\usepackage[utf8]{inputenc}
\usepackage[T1]{fontenc}
\usepackage{babel}
\usepackage{amsmath}
\usepackage{amsfonts}
\usepackage{csquotes}
\usepackage{amstext}
\usepackage{amssymb}
\usepackage{multicol}
\usepackage{graphicx}
\usepackage{comment}
\usepackage{subcaption}
\usepackage{fancyhdr}
\usepackage{siunitx}
\usepackage{hyperref}
\hypersetup{
    colorlinks=true,
    linkcolor=blue,
    filecolor=magenta,      
    urlcolor=cyan,
}
\usepackage{graphicx}

\usepackage[backend=biber,bibstyle=apa,style=nature,sorting=none]{biblatex}

\begin{document}


\title{Deep reinforcement learning for fMRI prediction of Autism Spectrum Disorder}

\author[1]{Joseph Stember\thanks{Corresponding author:- Joseph Stember - \href{mailto:joestember@gmail.com}{\texttt {joestember@gmail.com}}}}
\author[2]{Danielle Stember}
\author[1]{Luca Pasquini}
\author[1]{Jenabi Merhnaz}
\author[1]{Andrei Holodny}
\author[3]{Hrithwik Shalu}

\affil[1]{Memorial Sloan Kettering Cancer Center, New York, NY, US, 10065}
\affil[2]{New York University Langone Medical Center, New York, NY, US, 10016}
\affil[3]{Indian Institute of Technology, Madras, Chennai, India, 600036}
\makeatletter

\date{}

\maketitle
\thispagestyle{fancy}

\begin{abstract}

\indent \textit{Purpose} Because functional MRI (fMRI) data sets are in general small, we sought a data efficient approach to resting state fMRI classification of autism spectrum disorder (ASD) versus neurotypical (NT) controls. We hypothesized that a Deep Reinforcement Learning (DRL) classifier could learn effectively on a small fMRI training set. 

\indent \textit{Methods}  We trained a Deep Reinforcement Learning (DRL) classifier on 100 graph-label pairs from the Autism Brain Imaging Data Exchange (ABIDE) database. For comparison, we trained a Supervised Deep Learning (SDL) classifier on the same training set.

\indent \textit{Results} DRL significantly outperformed SDL, with a $\textit{p}$-value of $2.4 \times 10^{-7}$. DRL achieved superior results for a variety of classifier performance metrics, including an $F_1$ score of 76, versus 67 for SDL. Whereas SDL quickly overfit the training data, DRL learned in a progressive manner that generalized to the separate testing set.

\indent \textit{Conclusion} DRL can learn to classify ASD versus NT in a data efficient manner, doing so for a small training set. Future work will involve optimizing the neural network for data efficiency and applying the approach to other fMRI data sets, namely for brain cancer patients. 

\end{abstract}

\pagebreak

\section*{Abbreviations}

\begin{itemize}
    \item AI - Artificial Intelligence
    \item RL - Reinforcement Learning
    \item DRL - Deep Reinforcement Learning
    \item SDL - Supervised Deep Learning
    \item CNN - Convolutional Neural Network
    \item DQN - Deep Q Network
    \item GNN - Graph Neural Network
    \item fMRI - Functional MRI
    \item ASD - Autism Spectrum Disorder
    \item NT - Neurotypical
    \item TC - Neurotypical Controls
    \item ABIDE - Autism Brain Imaging Data Exchange
    \item TD(0) - Temporal Difference Zero
\end{itemize}

\section*{Funding sources}

We gratefully acknowledge support from the following sources:
\begin{itemize}
    \item American Society of Neuroradiology Research Grant in Artificial Intelligence
    \item Canon Medical Systems USA, Inc./Radiological Society of North America Research Seed Grant
    \item Memorial Sloan Kettering Cancer Center Radiology Developmental Project Fund
\end{itemize}

\section*{Conflicts of interest}

The authors have pursued a provisional patent based on the work described here.

\section*{Introduction}

Autism spectrum disorder (ASD) is a complex, heterogenous neurodevelopmental disorder characterized by atypical social communication and interactions as well as restricted or repetitive behaviors, interests, or activities \cite{american2013diagnostic}.  ASD often presents within the first two to three years of life, and it is a notably male-predominant disorder, with a prevalence four and half times higher among boys than girls. While estimates vary based on study methodology and populations, ASD prevalence in the United States in 2012 was estimated at one in 42 boys and one in 189 girls \cite{christensen2018prevalence}.

The cause of ASD remains elusive, with many potential genetic and environmental factors contributing. Genetic causes are supported by both twin and family heritability studies; among children born in Sweden, for instance, heritability of ASD was estimated at 50$\%$ with a ten-fold increase in the risk of autism if a full sibling has the diagnosis \cite{sandin2014familial}.  Genetic factors may alter brain development, with an abnormal gene affecting the expression of other genes (i.e. the "epigenetic theory"), or there may be interactions among multiple genes with environmental factors contributing \cite{lord2018autism}. Various environmental risk factors have been implicated in ASD as well, including advanced parental age, shortened interpregnancy interval, maternal hospitalization with infection during pregnancy, maternal diet, and possibly air pollution, among others \cite{lyall2017changing}.

In addition to genetic and environmental factors, abnormal neural connectivity has been implicated in the pathogenesis of ASD. There appear to be regions of hypoconnectivity, and, to a lesser degree, regions of hyperconnectivity, which has led to the hypothesis of ASD as a "disconnection syndrome" \cite{geschwind2007autism}. The Autism Brain Imaging Data Exchange I and II ("ABIDE" I and II) databases have been particularly invaluable in amassing a large data set and elucidating aberrant connectivity \cite{di2014autism,di2017enhancing}. Analyzing resting-state functional MRI (fMRI), structural MRI and phenotypical information from a combined 2,156 unique cross-sectional datasets has revealed how some areas of the brain demonstrate underconnectivity in ASD (particularly in cortico-cortical and interhemispheric circuits), with less frequent areas of aberrant overconnectivity (particularly in subcortico-cortical circuits) \cite{di2014autism,di2017enhancing}. Using this aggregate data and analyzing connectivity patterns and clinical profiles, several subtypes of ASD more recently have been described, with functional connectivity or lack thereof correlating with intelligence quotient \cite{reardon2021subtyping}. 

Despite the wealth of advances facilitated by the ABIDE I and II databases, the study of ASD using fMRI is limited by the need for large datasets \cite{liu2021autism}. Here we present a novel method of detecting ASD using small data sets with Deep Reinforcement Learning (DRL). The ABIDE I and II databases are a landmark, ground-breaking repository of information, and many discoveries have emerged from its being shared. By offering the additional tool of ASD detection employing smaller training sets, it is our hope that further ASD research can be applied to diverse, novel populations, and ASD subtypes, and that and even more can be learned about the pathophysiology and atypical neurocircuitry that underlies ASD.

DRL has shown early success in a wide variety of radiology artificial intelligence (AI) tasks, using various imaging modalities \cite{parekh2020multitask,al2019partial,alansary2019evaluating,maicas2017deep,ghesu2017multi,zhou2021deep,zhang2018deep,ali2018lung,jang2021deep,codari2020deep,winkel2020validation,winkel2021building,li2020deep,yin2021left,si2020multi,kooi2016comparison,xiong2021edge,zhang2021sequential,blair2018localization}. Recent DRL application to studying brain cancer on MRI has demonstrated DRL's remarkable ability to learn effectively from small training sets on these anatomic images \cite{stember2020deep,stember2020reinforcement,stember2020unsupervised,stember2021deep,stember2021deep_3D}. 

In the current work, we sought to show that DRL can also learn on small training sets of fMRI data. We hypothesized that DRL would learn from a small training set in a manner that generalizes to a separate testing set. We also hypothesized that Supervised Deep Learning would quickly overfit the small training set and hence not generalize well to the separate testing set. 

\section*{Methods}

\subsection*{Data collection and pre-processing}

We downloaded fMRI data from the publicly available and freely downloadable ABIDE database \cite{craddock2013neuro,di2014autism,di2017enhancing}. This data set contains resting state fMRI connectome data from 1,112 patients, with a breakdown of 539 individual with ASD, and 573 NT controls (TCs). 

We allotted 100 randomly selected data points as the training set, both for SDL and DRL. We set aside a separate set of 100 data points for the testing set. The components of the data for each subject of interest to us were the fMRI connectomes and corresponding labels. The latter indicated whether the subject had the clinical diagnosis of ASD versus NT. 

The training set contained 58 ASD cases and 42 TCs. The testing set contained 41 ASD cases and 59 TCs. 

\subsection*{Very brief overview of deep reinforcement learning (DRL)}

All reinforcement learning (RL) problems feature some environment in which we will train an agent. We define a policy as being the actions that our agent will take in any given situation (i.e. state). 

In DRL, the policy is given by a convolutional neural network (CNN) called the Deep Q Network (DQN). The DQN takes in the current state and calculates the probability of taking any of the permitted actions. Our ultimate task to to train the DQN such that the agent takes optimal actions to achieve our AI task (e.g. classification, segmentation or localization). 

The training process for the policy works as follows. Starting from a random initialization, we let the agent interact with the environment for a while and collect multi-step episodes of agent-environment interaction. This produces a recording of what sequence of states we encountered, what actions we took in each state, and what the reward was at each step. 

In our case, there are two possible actions: predicting that an fMRI is from an ASD versus NT subject. The states consist of the incidence matrix representing nodes (corresponding to brain regions) and edges, in addition to a scalar called $\text{pred}_\text{corr}$ that indicates whether the prior step's predicted action was correct ($\text{pred}_\text{corr}$=1). By default, $\text{pred}_\text{corr}$ is initially set to zero for each episode. This parameter provides the key to improving the policy; whatever the agent did leading up to the $\text{pred}_\text{corr}=1$ states was good, and whatever it did leading up to the $\text{pred}_\text{corr}=0$ states was bad. In other words, $\text{pred}_\text{corr}$ imbues the state with information about the state's value, information that ultimately transfers to the DQN parameters. 

Here, value loosely translates to average reward, which translates to how often the classifier makes correct predictions. By using temporal difference learning formulated in the two-step TD(0) version, we can sample the agent's interaction with the environment via the reward from its actions. We can then use backpropagation to compute a small update on the network’s parameters that would make correct predictions ($\text{pred}_\text{corr}=1$) more likely in the future, and the actions leading to incorrect predictions less likely in the future. As we do this, the DQN is also learning about the environment, due to the formulation of the Bellman Equation. A more formal description and formulation is available via the essential textbook on RL by Sutton and Barto \cite{sutton2018reinforcement}, and to a much more abbreviated degree is provided in Appendix A.

\subsection*{Deep-$Q$ Network (DQN) architecture}

Both SDL and DRL employ CNNs, which for the latter we refer to as DQNs. The graph structures that describe fMRI connectomes require a type of CNN called a Graph Neural Network (GNN). Specifically, we employed the BrainGNN developed and exemplified by Li, Zhou et al. \cite{li2020pooling,li2021braingnn}. We made use of both data pre-processing steps and the network architecture itself as available in the GitHub page associated with this publication \cite{li2020pooling,li2021braingnn}.

\begin{figure}[h!]
\centering
\includegraphics[width=12cm,height=7cm]{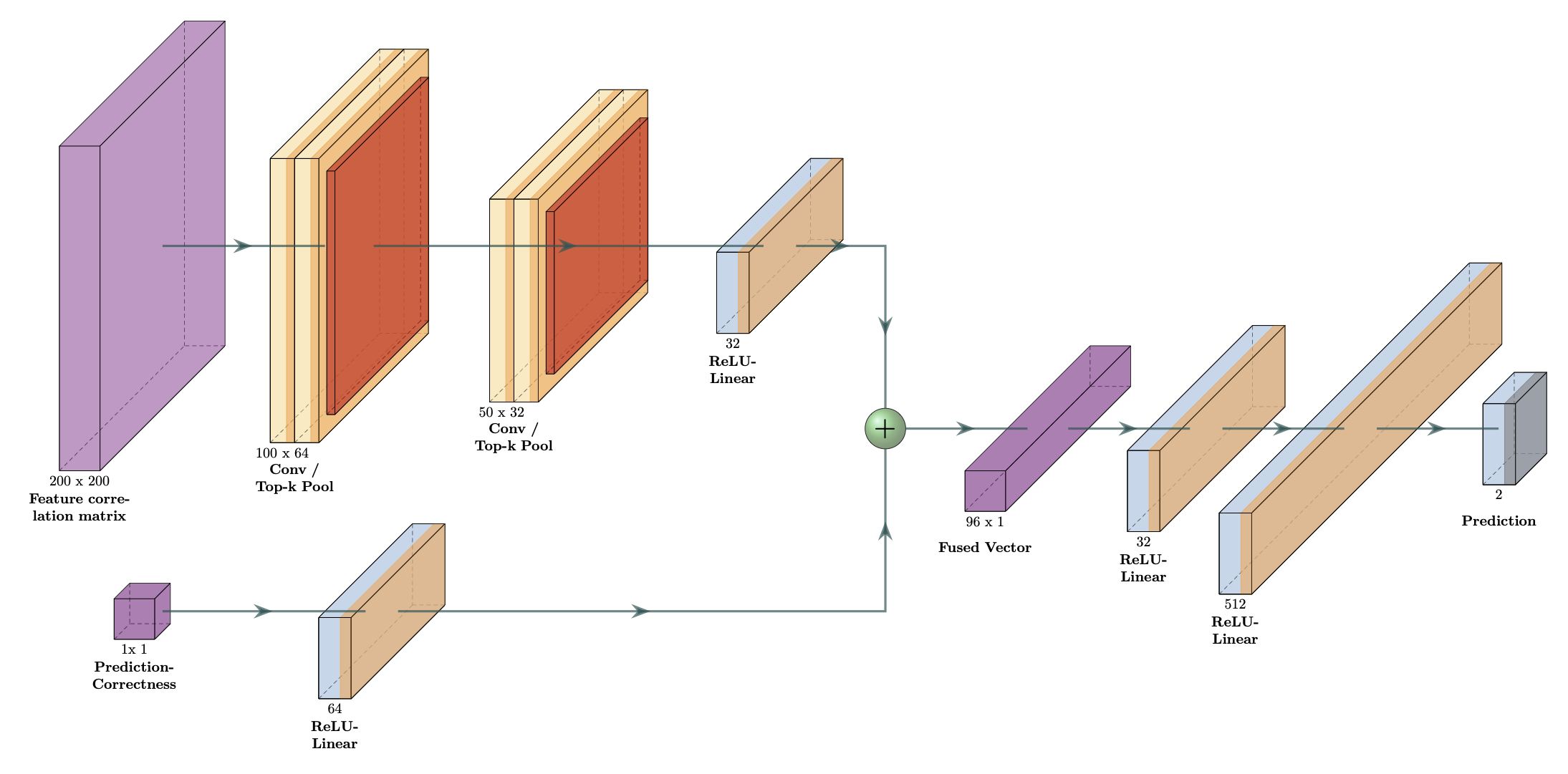}
\caption{ GraphGNN architecture with the addition of a scalar indicating prior step's prediction correctness for use in our Deep Q Network. }
\label{fig:DQN_architecture}
\end{figure}

A schematic illustrating the BrainGNN architecture is displayed in \textbf{Figure} \textbf{\ref{fig:DQN_architecture}}. The $200 \times 200$ incidence matrix describing each subjects brain regional edges (rows) and nodes (columns) serves as input to one branch of the network. This first/input layer represents the fine-grained functional connectivity, essentially a most "zoomed in" view of the connectome. It undergoes convolutions with rectangular kernels of length equals to the rows or columns, so as to maintain the full information regarding nodes and edges. Convolution occurs first over the edges before the nodes. There are more likely to be redundant edges than redundant nodes, and the order of operations permits the training process to "weed out" redundant edges first via the pooling layers. Following the convolutions, the resulting layers are passed through a type of pooling called top k pooling, essentially meant to remove redundant or noncontributory edges, i.e. edges that do not contribute to the classification results.  

The scalar $\text{pred}_\text{corr}$ enters the GraphGNN via a separate pathway, ultimately concatenated to the flattened output of the convolution-pooling branch. This concatenated/fused layer then proceeds to a fully connected layers of 512, 128 then 64 neurons before finally reaching the 2-neuron output.

Additional details of network implementation are provided in Appendix B. We performed all calculations using Google Colab Pro, employing their Tesla T4 GPU and Intel(R) Xeon(R) CPU with 2.00GHz processor. We performed the implementation on PyTorch with CUDA backend. 

\subsubsection*{Supervised deep learning (SDL) classification for comparison}

For comparison to DRL, we trained an SDL-based classifier on the same training set. We used similar GraphGNN architecture to that of the DQN shown in \textbf{Figure} \textbf{\ref{fig:DQN_architecture}}. However, we only used one branch, the convolution-pooling branch. Our input was the incidence matrix. We trained for 100 epochs, using negative log likelihood cross entropy loss, with Adam optimization and learning rate of $1 \times 10^{-2}$. Additionally, as per the SDL GraphGNN approach of of Li, Zhou et al. \cite{li2021braingnn,li2020pooling}, we employed batch normalization, dropout, with additional regularization provided by top k loss and consistency loss terms \cite{li2021braingnn,li2020pooling}.

\section*{Results}

\begin{table}[h!]
\centering
\begin{tabular}{ |p{4.2cm}|p{1.6cm}|p{1.6cm}|  }
\hline
\multicolumn{3}{|c|}{Comparison of DRL and SDL} \\
\hline
\textbf{Quantity} & \textbf{DRL} (\%) & \textbf{SDL} (\%) \\ 
\hline
Sensitivity & 100 & 100  \\
Specificity & 39 & 0 \\
Positive predictive value & 62 & 50 \\
Negative predictive value & 100 & NaN  \\
$F_1$ score & 76 & 67 \\
\hline
\end{tabular}
\caption{Comparison for several model performance quantities between Supervised Deep Learning (SDL) and Deep Reinforcement Learning (DRL). The last quantity, the $F_1$ score, is the harmonic mean of precision and specificity}
\label{table:comp_table}
\end{table}

\begin{figure}[h!]
\centering
\includegraphics[width=11.5cm,height=6.5cm]{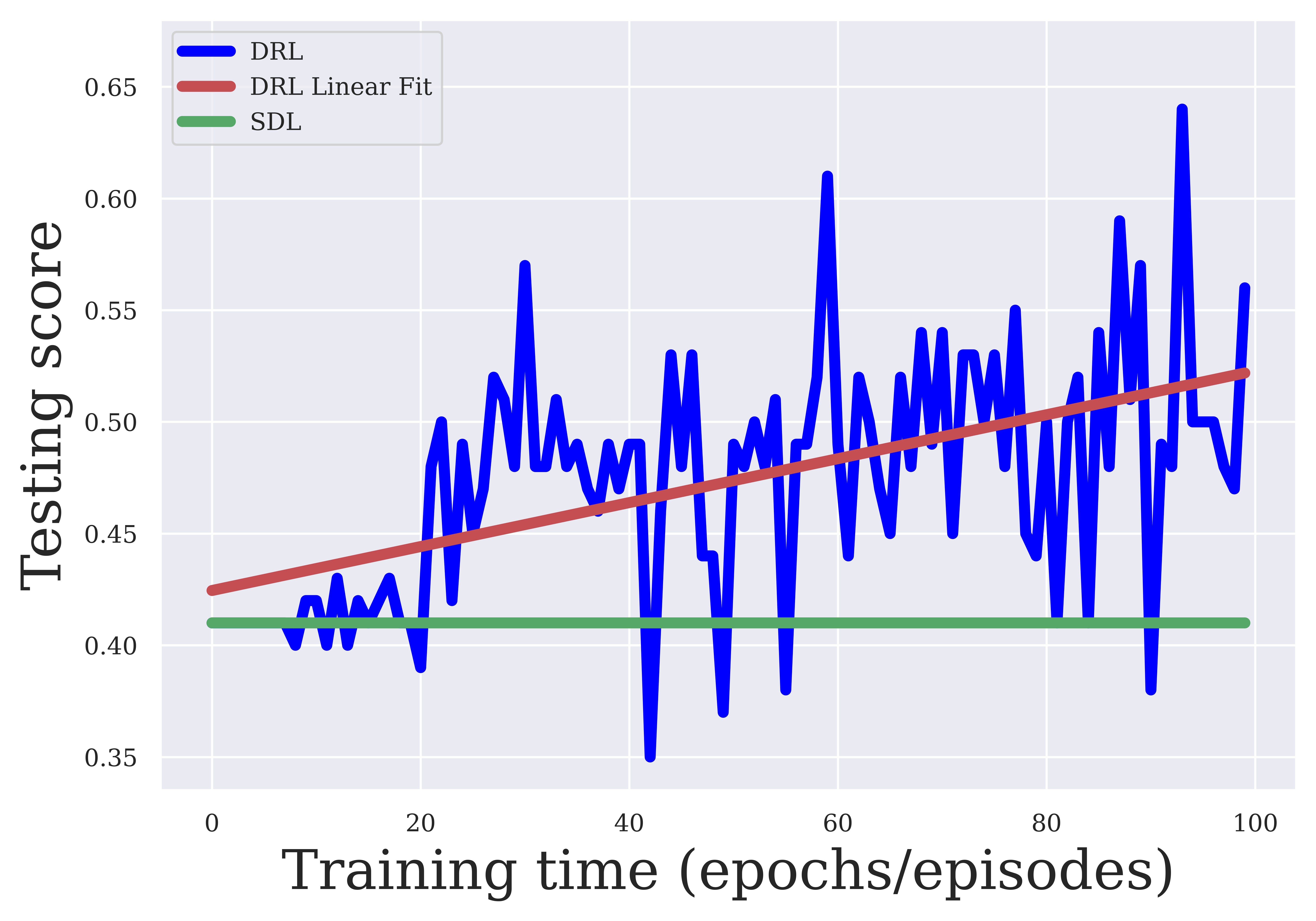}
\caption{ DRL (blue) versus SDL (green) testing set results. The best fit line for the DRL data is shown in red, with a slope of 0.001. Of note, episode increments actually represent every tenth episode during training.}
\label{fig:test_results}
\end{figure}

As we have seen for anatomic imaging small data AI, SDL quickly overfits the small training set. We again see that, in contrast, DRL is able to learn progressively on the small training set. This is manifested by overall improving performance on the separate testing set in the case of DRL, as displayed in \textbf{Figure} \textbf{\ref{fig:test_results}}. 

\begin{figure*}[h!]
\begin{multicols}{2}
    \includegraphics[width=\linewidth]{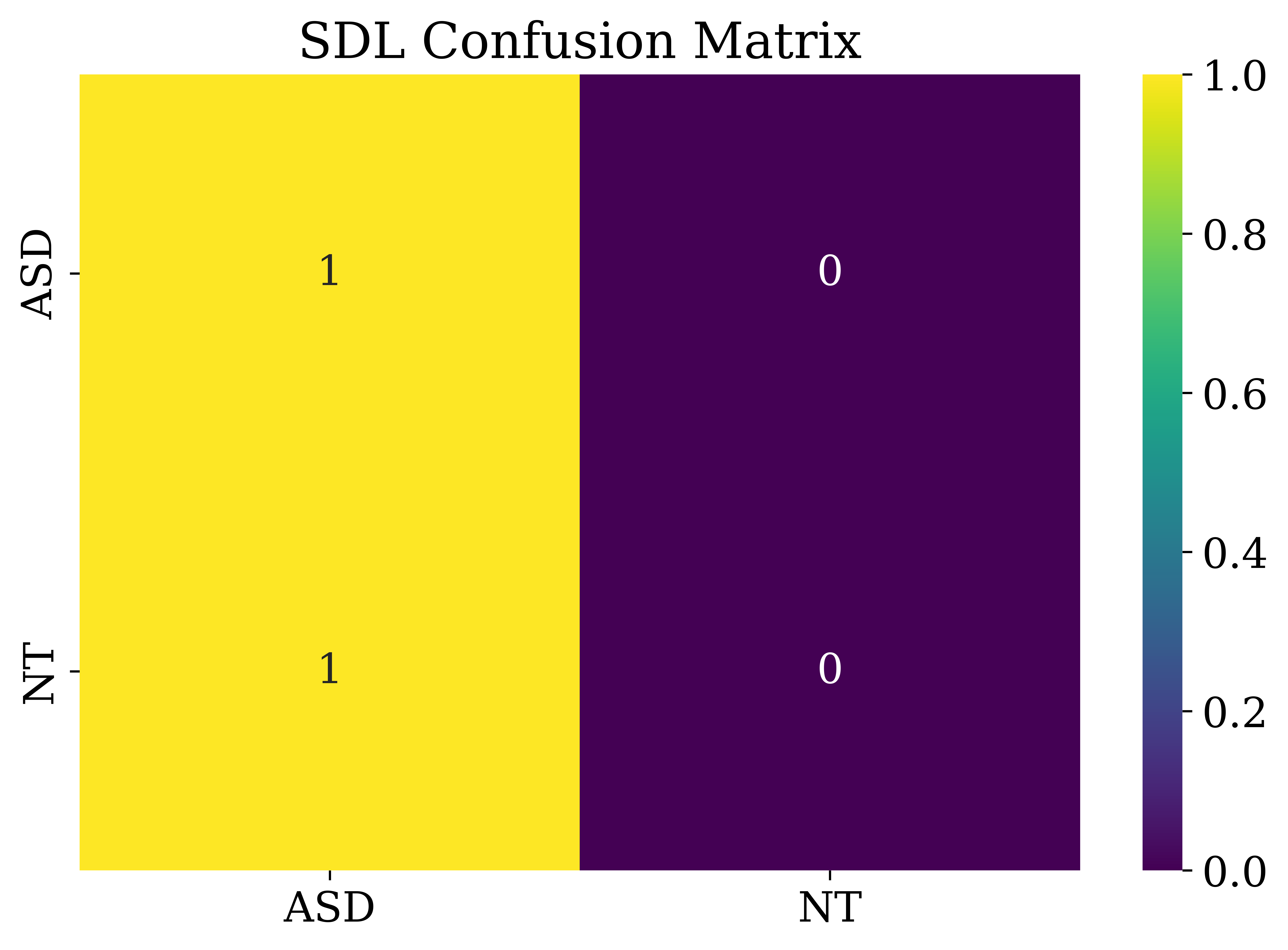}\par 
    \includegraphics[width=\linewidth]{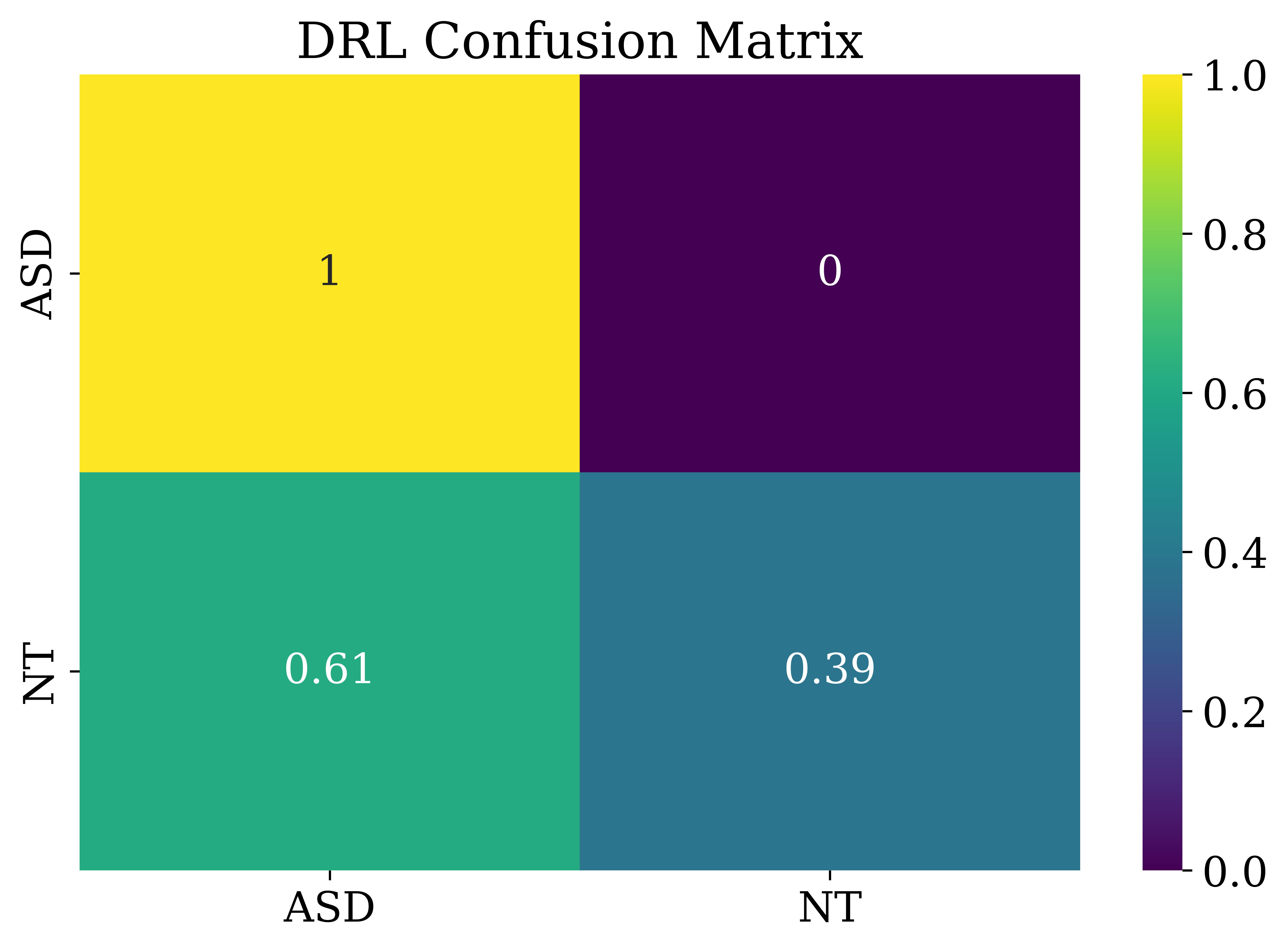}\par 
    \end{multicols}
\caption{Confusion matrices for Supervised Deep Learning (SDL) and Deep Reinforcement Learning (DRL). These are both scaled for maximum value of one.}
\label{fig:conf_mtx}
\end{figure*}

This figure displays that SDL does not learn in a way that generalizes to the separate testing set; the slope of its testing set accuracy is zero. The training set has a majority class of NT subjects (NT:ASD ratio of 58:42). This combines with the small training set size to bias SDL toward classifying fMRI data as NT. As such, SDL predicts only NT for the testing set, and misses all of the ASD cases. In contrast to the training set, the testing set is majority ASD patients (59 subjects), with a minority of 41 NT subjects. SDL quickly converges to 41$\%$ accuracy by predicting NT in all cases. 

In contrast, DRL learns in a manner that generalizes to the testing set. A best fit line through DRL testing set values shows the trend of increasing accuracy continued learning over time. The line has a slope of $+0.001$. The peak values within the DRL results are the optimal policies, and we can see these increase as well, one reach a maximum of 64$\%$ accuracy. The variations, or noise, in the DRL testing set accuracy reflect the off-policy search process that uncovers new optimal policies.

The confusion matrices for DRL and SDL are juxtaposed in \textbf{Figure} \textbf{\ref{fig:conf_mtx}}. From these matrices, we were able to compute a variety of quantities that assess model performance. Namely, we calculated the following: Sensitivity, specificity, positive predictive value, negative predictive value and $F_1$ score. The results are displayed in \textbf{Table} \textbf{\ref{table:comp_table}}. The table shows superior performance across metrics by DRL. 
 
Furthermore, by McNemar's test, we found statistically significantly improved classification performance of DRL over SDL, with a $\textit{p}$-value of $2.4 \times 10^{-7}$.

\section*{Discussion}

Data efficiency is an important goal in AI for fMRI analysis. We need data efficient algorithms because fMRI data sets are often inherently small, given the time-consuming, costly process of acquisition and extensive post-processing required, even at academic institutions. Additionally, if we wish to classify rare diseases or subclasses of conditions, such as those of ASD, the training sets become divided and smaller still.

To strive for data efficiency, we have demonstrated an application of DRL for binary classification between ASD and NT on a small fMRI training set. Prior studies have demonstrated DRL's ability to learn based on very small training sets of anatomic imaging-based data. We have shown a similar ability for fMRI graph-based data. 

A limitation is that one of the branches of our DQN used the particular GraphGNN formulation that was built principally for SDL and interpretability \cite{li2020pooling,li2021braingnn}. We suspect that a wholly custom-built architecture designed for data efficiency would allow DRL to classify with higher accuracy. 

In future work, we plan to apply DRL to fMRI data sets of brain cancer patients, both primary glioma/glioblastoma and metastatic brain disease.  
\section*{Conclusion} We have shown that DRL outperforms SDL for predicting ASD based on small fMRI training sets. Going forward, we hope that the ability to train effectively on small fMRI training sets will form an important component of personalized medicine. 

\section*{Appendix/Supplementary Materials}

\appendix

\section{Reinforcement Learning}

\subsubsection*{Reinforcement learning environment, definition of states, actions, and rewards}

As in all RL formulations, we must define our environment, itself defined by the agent, possible states, actions, and rewards.

\subsubsection*{Agent}

The agent in any RL formulation is an intuitive but formally murky concept. Essentially, the agent has a policy associated with it according to which the agent performs actions in various states and receives certain rewards. The goal of reinforcement learning here is to optimize the agent so as to act according to a policy (prescribed set of actions) that maximizes the total expected discounted reward received. Hence, our goal, which aligns with that of the agent, is to learn the optimal policy for the agent to follow. The problem formulation can be thought of as learning an optimal non-linear policy in a finite state discounted Markov Decision Process. 

\subsubsection*{States}

Here, the state is defined by an fMRI feature matrix plus a scalar quantity that indicates whether the previous class prediction (NT versus ASD) was correct. We denote this prediction correctness by $\text{pred}_{\text{corr}}$. It is defined by 

\begin{equation}\label{corr_pred_defn_1}
    \text{pred}_{\text{corr}} = \begin{cases}
        1 \text{, if prediction is correct} \\
        0 \text{, if prediction is wrong}  \text{.}
    \end{cases}
\end{equation}

Hence, we can write the state $s$ more formally as 
\begin{equation}\label{state_formal}
    s = \{ \mathcal{M}, \text{pred}_{\text{corr}} \},
\end{equation}
where $\mathcal{M}$ is the $200 \times 200$- matrix representing the fMRI data.

\subsubsection*{Actions}

The two possible actions, $a_1$ and $a_2$, are simply the prediction of whether the state $s$ corresponds to NT or ASD; i.e., the actions $\mathcal{A}$ are given by,
\begin{equation}
    \mathcal{A} = \begin{pmatrix} a_1 \\ a_2 \end{pmatrix} = \begin{pmatrix} 0 \\ 1 \end{pmatrix} = \begin{pmatrix} \text{predict NT} \\ \text{predict ASD} \end{pmatrix} \text{.}
\label{action_eqn}
\end{equation}

\subsubsection*{Rewards}

In order to encourage correct class predictions, we employ a reward system that incentivizes correct predictions. Hence, we provide a reward of $+1$ if the prediction is correct and penalize with reward of $-1$ for a wrong prediction. We can define the reward $r$ in terms of prediction correctness, $\text{pred}_\text{corr}$:
    \begin{equation}\label{rewards}
    r = \begin{cases}
        -1 \text{, if $\text{pred}_\text{corr} = 0$} \\
        +1 \text{, if $\text{pred}_\text{corr} = 1$}  \text{.}
    \end{cases}
\end{equation}

\subsubsection*{Action-value function ($Q$)}

A fundamentally important quantity in RL is that of a value function, in our case taking the form of the action-value function, denoted by $Q(s,a)$. $Q(s,a)$ essentially tells us the expected total cumulative reward that the agent would receive by selecting a particular action given a particular state, then acting in an optimal (on-policy) manner afterwards until the end of the current episode. An episode is defined as a number of states after which the agent begins anew in a different initial state. Restated, the action-value function represents the expected cumulative reward for taking action $a$ while in state $s$, if upon taking action $a$, a given policy (set protocol for selecting actions) is pursued thereafter until the end of the episode. More formally, given a policy $\pi$, the corresponding action-value, $Q^{\pi}(s,a)$, is defined by:
\begin{equation}\label{Q_defn}
    Q^{\pi}(s,a)=E_{\pi}\{R_t \arrowvert s_t=s,a_t=a\}=E_{\pi}\{\sum_{k=0}^{\infty} \gamma^k r_{t+k+1} \arrowvert s_t=s,a_t=a \} \text{,}
\end{equation}
where $R_t$ is the total cumulative reward starting at time $t$ and $E_{\pi}\{R_t \arrowvert s_t=s,a_t=a\}$ is the expectation for $R_t$ upon selecting action $a$ in state $s$ and subsequently picking actions according to $\pi$. The discount factor $\gamma$ represents the trade-off between weighting immediate rewards ("instant gratification") with rewards later on ("delayed gratification"). We set it equal to 0.99, with scaled reduction per episodes so as to maximise initial rate of sampling.

The action-value function is critically important, because by maximizing this quantity, we can ultimately reach an optimal policy that produces a desired behavior, in this case correctly predicting image class.

\subsubsection*{Deep-$Q$ Network}

In order to predict the actions our agent will take, we use the Deep-$Q$ Network (DQN) as a typical non-linear function approximator.

Our DQN consists of two arms / branches that join toward the end. In one branch, the fMRI matrix serves as input, undergoing graph convolutions.

In a separate, parallel pathway, $\text{pred}_{\text{corr}}$ is passed to a flattened layer, which is then concatenated with the last fully connected layer of the fMRI martix network branch. This concatenated layer is then connected multilayer perceptron with a two-node output. 

The two-node output represents the two possible action-values, $Q(s,a_1)$ and $Q(s,a_2)$, that result from taking actions $a_1$ and $a_2$, respectively, from state $s$.  Again, since we wish to maximize the total cumulative reward, we should maximize $Q(s,a)$. We do so simply by selecting the $\text{argmax}_a \left( Q \right)$, thereby selecting the action that maximizes expected cumulative reward. 

We should note at this point that selecting $\text{argmax}_a \left( Q \right)$ is an "on-policy" action selection. We need to experiment with random action selections to explore and learn about the environment. This is called off-policy behavior. 

Indeed, an initial conundrum / catch-22 is that we wish to train our network to approximate the optimal policy's $Q$ function, but at first we have no idea what that optimal policy is. We can only start to learn it in pieces by sampling from the environment via off-policy exploration and Temporal Difference Learning. 

\subsubsection*{Temporal Difference Learning}

While the DQN allows us to select actions for our agent to take, we need to learn the best policy via the process of taking actions and receiving rewards. The rewards in particular tell us about our environment. We have to store these "experiences" of the agent in order to better understand the environment. By doing so in tandem with the use of DQN, we can train the parameters of the DQN to build it into a reliable approximator for the action-value function $Q(s,a)$. However, the optimal policy, and thus best possible action-value function, is a moving target that we approach through sampling. 

Then, using the DQN $Q$ function approximator for an ever-improving policy's $Q$ function, we can better explore and increasingly exploit what our agent knows about the environment to sample the environment more efficiently. The end result of this process is that DQN becomes a better and better approximator for not only a $Q(s,a)$ function, but ultimately in theory the "global" optimal $Q(s,a)$ function, denoted by $Q^{\star}$. In practice, the exploration continues throughout our training, since the goal is to find "local" optimal policies that produce high testing set accuracy.

In general, an episode of training proceeds as follows:

We select one of the training set images at random. The initial predicted class (normal or tumor-containing) is guessed at random. The initial value for $\text{pred}_\text{corr}$ is set to zero as a default. At this point, an action is taken, i.e., a prediction as to the image class is made. As before, this is done in a manner that initially is more random, in order to emphasize exploration of the environment. As the agent learns about the environment and the best policy to follow, the degree of randomness decreases and the agent increasingly chooses the optimal action predicted by the DQN, i.e., the $\text{argmax}_a\{Q(a=0),Q(a=1)\}$. This is called the epsilon-greedy algorithm.
\begin{itemize}
    \item $\epsilon = 0.7$
    \item $\epsilon_{min} = 1 \times 10^{-4}$
    \item $\Delta \epsilon = 1 \times 10^{-4}$,
\end{itemize}
where $\epsilon$ is the initial exploration rate, which decreases during training at a rate of $\Delta \epsilon$ down to a minimum value of $\epsilon_{min}$. Again, although RL in theory always converges toward a global optimal policy, this is often not practically feasible. However, that is acceptable given our goal of identifying "locally" optimal policies that produce high-accuracy testing set predictions. In fact, the duration of the training of 1,000 episodes used here results in a final $\epsilon = 0.5$. As such, even at the end of our training, the agent performs random environmental exploration as much as exploitation.

\section{RL Training: hyperparameters}

\begin{itemize}
    \item Loss: mean squared error
    \item Adam optimizer
    \item learning rate: $1 \times 10^{-2}$
    \item Batch size: 32
    \item Training time: 1,000 episodes
    \item 5 steps per episode
    \item Replay Memory Buffer: 20000 transitions 
\end{itemize}

\printbibliography

\end{document}